\documentclass[%
reprint,
amsmath,
amssymb,
aps,
prb,
floatfix,
superscriptaddress,
longbibliography,
]{revtex4-2}

\usepackage{hyperref}
\usepackage[capitalise]{cleveref}

\usepackage{makecell}
\usepackage{graphicx}
\usepackage[utf8]{inputenc}
\usepackage{dcolumn}
\usepackage{bm}
\usepackage[colorinlistoftodos]{todonotes}
\presetkeys{todonotes}{inline}{}
\usepackage{chemformula}
\usepackage{xspace}
\usepackage{mathtools}
\usepackage{siunitx}
\usepackage{notes2bib}
\usepackage{multirow}

\newcommand{\Fe}{\ch{PbFeBO4}\xspace}
\newcommand{\Mn}{\ch{PbMnBO4}\xspace}
\newcommand{\Cr}{\ch{PbCrBO4}\xspace}
\newcommand{\Fam}{\ch{PbMBO4} (\ch{M}=\ch{Cr}, \ch{Mn}, \ch{Fe})\xspace}

\newcommand{\abin}{\textit{ab initio}\xspace}

\newcommand{\cm}{cm$^{-1}$\xspace}
\newcommand{\TN}{$T_N$\xspace}

\begin{document}

\title{One- and two-magnon excitations in antiferromagnet \Fe}

\author{M. A. Prosnikov}
 \email{yotungh@gmail.com}
 \affiliation{High Field Magnet Laboratory (HFML - EMFL), Radboud University, Toernooiveld 7, 6525 ED Nijmegen, The Netherlands}
 \affiliation{Radboud University, Institute for Molecules and Materials, Heyendaalseweg 135, 6525 AJ Nijmegen, The Netherlands}
\date{\today}

\begin{abstract}
The linear spin wave theory study of \Fe spin dynamics is presented.
It is shown that the modes observed in Raman scattering experiments below N\'eel temperature in~\cite{prosnikov_pbfe_2016} are optical magnon and two-magnon excitations.
Based on the magnon energy, two-magnon band lineshape, and Weiss temperature~\cite{pankrats_pbfe_2014}, the consistent set of the exchange coupling constants up the third neighbor is derived and compared with the results of \abin calculations~\cite{koo_density_2009,xiang_intra-chain_2016,curti_PbFe_2019}.
The small deviation of the observed two-magnon band from the one-magnon density of states suggests a surprisingly negligible role of magnon-magnon interactions.
\end{abstract}


\maketitle

\section{\label{sec:intro}Introduction}

The promising field of antiferromagnetic spintronics~\cite{nemec_antiferromagnetic_2018,gomonay_spintronics_2014,baltz_antiferromagnetic_2018,jungwirth_antiferromagnetic_2016} constantly demands the discovery of the new functional materials with specified properties and the development of reliable theoretical models.
Some potential material candidates manifest intrinsic coupling of different subsystems such as magnetic, orbital, electronic, and lattice, allowing for additional degrees of freedom to control spin excitations~\cite{spaldin_advances_2019,son_unconventional_2019,han_lattice_2019}. 

The \Fam family of compounds belongs to the sillimanite group \cite{fischer_crystal_2008}, where the presence of the stereochemically active Pb$^{2+}$ cations leads to the reduction of the connectivity between magnetic ions resulting in the unique topology of the exchange structure~\cite{murshed_role_2012}.
Moreover, the types of magnetic ions drastically affect magnetic properties, such as magnetic structures, critical temperatures, and dispersion of the magnetic excitations without change of crystal symmetry.
Notably, \Mn is an extremely rare example of insulating ferromagnets~\cite{park_synthesis_2003,pankrats_pbmn_2016}, while other (\Fe and \Cr) are known to be antiferromagnets~\cite{park_synthesis_2003}.
There are a few predicted compounds with other 3$d$ ions \ch{PbMBO4~(M = Ti, V, Co)}~\cite{xiang_intra-chain_2016}, which yet to be synthesized.

In this family, the \Fe exhibits the highest transition temperature of \TN$ = 114$\,K, shows anisotropic and negative thermal expansion observed with X-ray and neutron diffraction~\cite{murshed_anisotropic_2014}, anomalies in the vicinity of \TN in both dielectric susceptibility~\cite{pankrats_pbfe_2014} and phonon energies~\cite{prosnikov_pbfe_2016} indicating coupling between magnetic and lattice subsystems.
Magnetostatic and dielectric properties of \Fe and \Mn were studied in detail in~\cite{pankrats_pbfe_2014, pankrats_pbmn_2016}.
There are a number of \textit{ab initio} calculations~\cite{koo_density_2009,xiang_intra-chain_2016,curti_PbFe_2019,curti_SnM_2019} dedicated to the determination of the exchange constants.
Simultaneously, the reliable determination of the exchange constants is the crucial step in understanding both the static and dynamical properties of the material and its further potential for applications.

In this paper, we report on the linear spin-wave theory calculations allowing us to derive closed-form magnon dispersion relation for \Fe (and equivalent compounds), calculation of the two-magnon (2M) band lineshape, and ground-state phase diagram for exchange couplings up to the third neighbor.
The consistent set of the exchange constants ($\text{J}_0, \text{J}_1,\text{J}_2$) is proposed based on the experimentally observed energy of the optical branch, shape of 2M band~\cite{prosnikov_pbfe_2016}, and Curie-Weiss temperature~\cite{pankrats_pbfe_2014}.
It is shown that both interchain couplings ($\text{J}_1, \text{J}_2$) are crucial to capture peculiarities of the spin dynamics.
Their values, considering coordination numbers, are comparable with intrachain one the ($\text{J}_0$)classifying \Fe as 3D Heisenberg antiferromagnet.
The symmetry allowed Dzyaloshinskii-Moriya interaction (DMI) on $\text{J}_0$ path could explain magnetic susceptibility anomaly~\cite{pankrats_pbfe_2014} in the absence of weak ferromagnetic moment and could be directly observed by the zero-field splitting of the acoustical magnon branch.
The estimated energy range of magnetic excitations for \Cr is briefly discussed at the end.

\section{Results and discussion}\label{sec:results}

\begin{figure*}
    \centering\includegraphics[width=14cm]{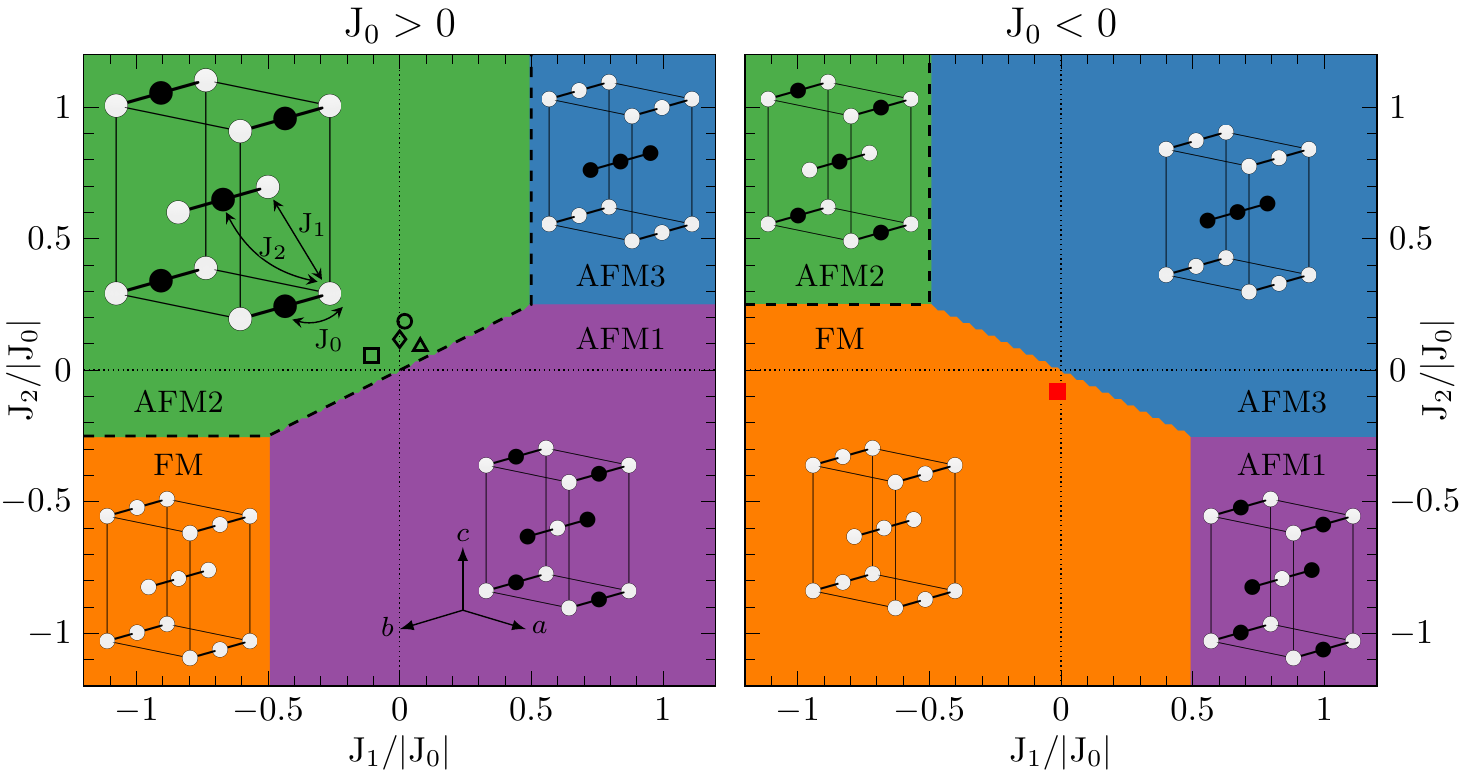}
    \caption{
        \label{fig:phase_diagram}
        Ground state magnetic phase diagrams as a function of interchain exchange interactions J$_1$ and J$_2$ normalized at intrachain one J$_0$ obtained through energy minimization (colored regions) and according to the real domain of \cref{eq:sw_modes} (thick dashed line).
        Left and right panels correspond to antiferromagnetic and ferromagnetic intrachain interaction, respectively.
        Insets depict structures with only magnetic ions shown.
        Marks shows sets of exchange parameters for \Fe calculated in this work (square), Koo et al.~\cite{koo_density_2009} (circle), Xiong et al.~\cite{xiang_intra-chain_2016} (triangle), Curti et al.~(set b)~\cite{curti_PbFe_2019} (diamond), and for \Mn Koo et al.~\cite{koo_density_2009} (red square).
    }
\end{figure*}

Below \TN$ = 114$\,K \Fe undergoes paramagnetic to antiferromagnetic phase transition with magnetic propagation vector $\bm{k} = 0$ according to neutron diffraction measurements~\cite{park_synthesis_2003}.
The resulting magnetic structure can be described as coupled antiferromagnetic chains of \ch{[FeO6]} octahedra.
Based on exchange interactions up to the third neighbor, the $\bm{k} = 0$ ground state phase diagram is calculated through energy minimization~\cite{sandor_toth_2017_838034} and shown in~\cref{fig:phase_diagram}.
Two cases of antiferromagnetic and ferromagnetic intrachain interactions were considered, and it is shown that all possible $\bm{k} = 0$ magnetic structures could be realized in both cases, however taking into account dominant role of J$_0$ the most probable structures are AFM2 and AFM1 for J$_0 > 0$ and FM and AFM3 for J$_0 < 0$, respectively.

\begin{figure*}
    \centering\includegraphics[width=17.8cm]{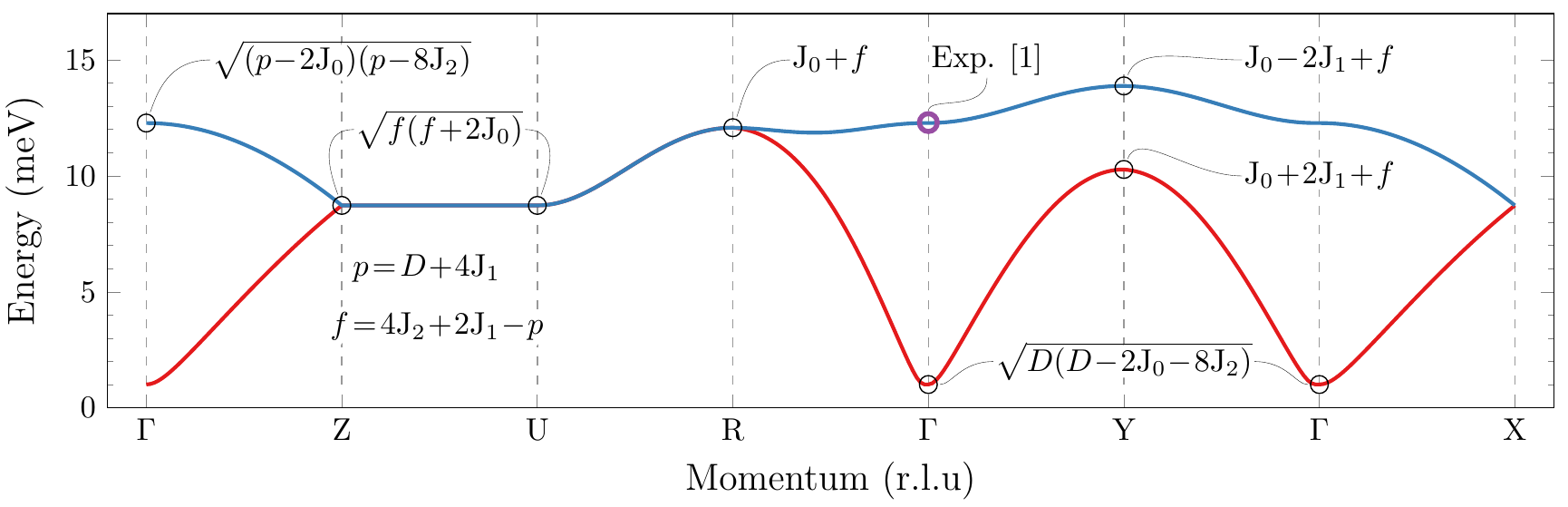}
    \caption{
        \label{fig:PbFe_dispersion}
        Spin-waves dispersion along the high-symmetry path in the Brillouin zone according to~\cref{eq:sw_modes}.
        To obtain the energy of the highlighted points equations should be multiplied by the $2S$ (e.g. $S=5/2$ for \Fe).
        Purple circle shows the energy of the experimentally observed optical branch~\cite{prosnikov_pbfe_2016}.
    }
\end{figure*}

The following Hamiltonian based on isotropic exchange interactions and single-ion anisotropy (SIA) terms is considered for spin-wave calculations:

\begin{equation} \label{eq:PbFe_Hamiltonian}
    \begin{gathered}
        \mathcal{H} = \sum_{\mathclap{\langle i,j \rangle}} \text{J}_0 \mathbf{S}_i \mathbf{S}_j + \sum_{\mathclap{\langle \langle i,j \rangle \rangle}} \text{J}_1 \mathbf{S}_i \mathbf{S}_j + \sum_{\mathclap{\langle \langle \langle i,j \rangle \rangle \rangle}} \text{J}_2 \mathbf{S}_i \mathbf{S}_j + \sum_{i} \text{D} ({\mathbf{S}_i^z})^2\ ,
    \end{gathered}
\end{equation}

where $\mathbf{S}$ is the spin operator, J$_0$..J$_2$ stands for superexchange constants corresponding to paths shown in~\cref{fig:phase_diagram}.
$\text{J}>0$ corresponds to AFM exchange.

Utilizing Holstein-Primakoff transformations~\cite{holstein_field_1940} and Fourier transformation of the exchange couplings the bosonic matrix form of the Hamiltonian is obtained:

\begin{equation} \label{eq:ham_matrix}
    \begin{aligned}
            \left(
                \begin{array}{cccccccc}
                     a & b^* & 0 & 0 & 0 & 0 & c^* & e^* \\
                     b & a & 0 & 0 & 0 & 0 & d & c^* \\
                     0 & 0 & a & b^* & c & d^* & 0 & 0 \\
                     0 & 0 & b & a & e & c & 0 & 0 \\
                     0 & 0 & c^* & e^* & a & b^* & 0 & 0 \\
                     0 & 0 & d & c^* & b & a & 0 & 0 \\
                     c & d^* & 0 & 0 & 0 & 0 & a & b^* \\
                     e & c & 0 & 0 & 0 & 0 & b & a \\
                \end{array}
            \right)\ ,
    \end{aligned}
\end{equation}

where

\begin{equation} \label{eq:ham_elements}
    \begin{aligned}
        a &= 2 \text{S}~(-\text{D}+\text{J}_0 - 2 \text{J}_1 + 4 \text{J}_2) ,\\
        b &= \text{S}~\text{J}_1 \left(1 + e^{2 \pi i h}\right)  \left(1 + e^{2 \pi i l}\right) ,\\
        c &= \text{S}~\text{J}_0 \left(1 + e^{2 \pi i k}\right) ,\\
        d &= 8 \text{S}~\text{J}_2  \cos (\pi h) \cos (\pi k) \cos (\pi l) e^{\pi i (h-k+l)} ,\\
        e &= \text{S}~\text{J}_2 \left(1 + e^{2 \pi i h}\right)  \left(1 + e^{2 \pi i k}\right) \left(1 + e^{2 \pi i l}\right)\ .
    \end{aligned}
\end{equation}

\begin{figure*}
    \centering\includegraphics[width=14cm]{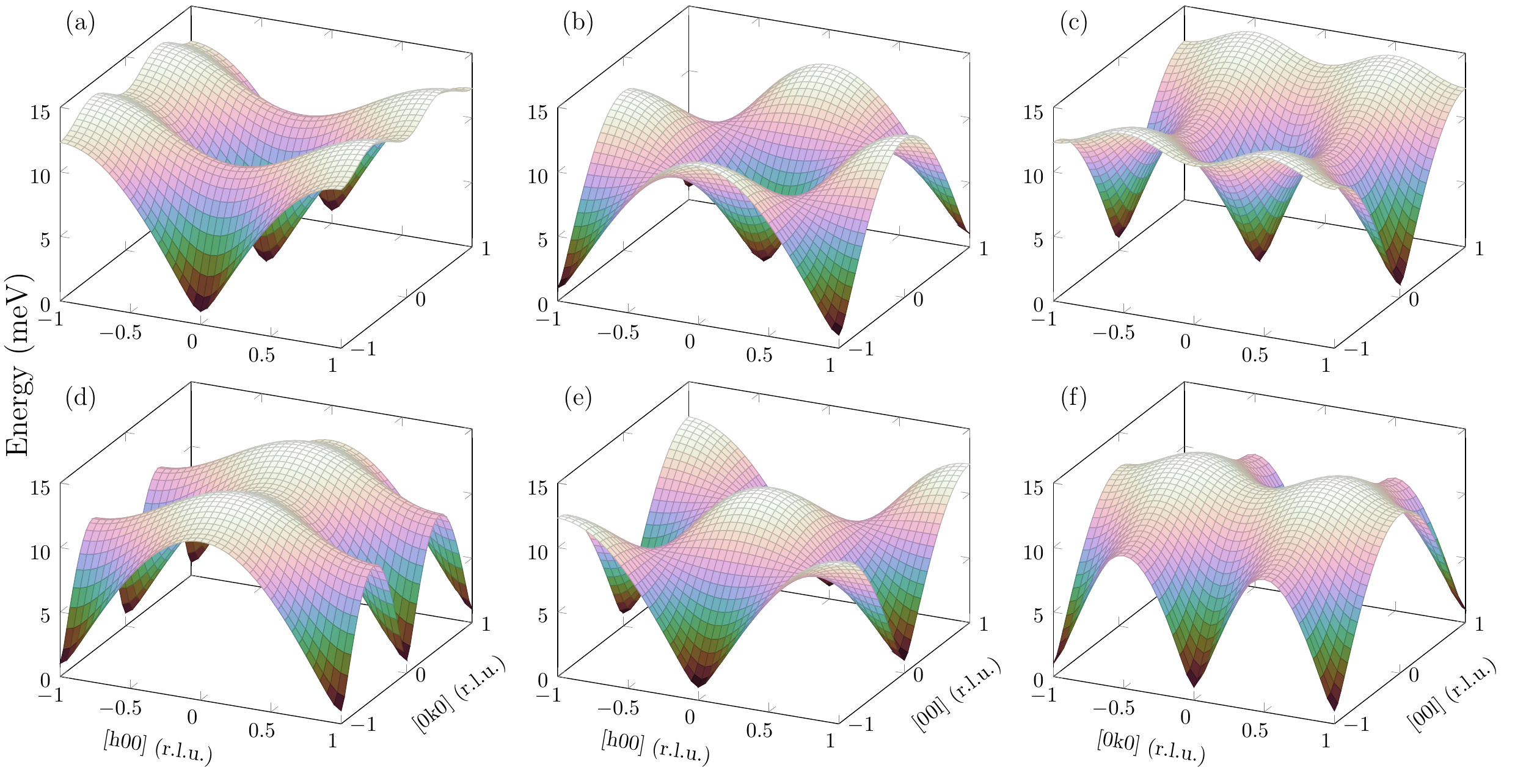}
    \caption{
        \label{fig:PbFe_dispersion_3D}
        Spin-wave dispersion for \Fe given by \cref{eq:sw_modes} with exchange coupling constants from the first row of~\cref{table:constants}. 
    }
\end{figure*}

Diagonalization~\cite{white_diagonalization_1965} of the matrix~\cref{eq:ham_matrix} leads to two positive doubly-degenerate, in the absence of external magnetic field, spin-wave modes corresponding to acoustical and optical branches:  


\begin{equation} \label{eq:sw_modes}
    \begin{aligned}
        \omega & = 2 \text{S} \big[
        (\text{D}-\text{J}_0+2 \text{J}_1-4 \text{J}_2)^2 \\
        & \mp 4 \cos(\pi h) \cos(\pi l) (\text{D} \text{J}_1 + 2 \text{J}_1 (\text{J}_1 - 2 \text{J}_2) \\
        & + \text{J}_0 (\text{J}_2 - \text{J}_1) + \text{J}_0 \text{J}_2 \cos(2 \pi k) \mp \text{J}_1^2 \cos(\pi h) \cos(\pi l)) \\
        & - \cos^2(\pi k) (\text{J}_0^2 + 16 \text{J}_2^2 \cos^2(\pi h) \cos^2(\pi l))
        \big]^{1/2}\ ,
    \end{aligned}
\end{equation}

where $h,k,l$ are given in reciprocal lattice units, and different branches are distinguished by $\mp$ sign.
Obtained dispersion curves and surfaces are shown in~\cref{fig:PbFe_dispersion} and~\cref{fig:PbFe_dispersion_3D}, respectively.

All the calculations and plots were done with a small value of the single ion easy-axis type anisotropy of $-0.01$~meV (along the $c$ axis) to reproduce the observed magnetic structure of \Fe~\cite{park_synthesis_2003}.
Despite the fact that it is impossible to derive the precise value of SIA based on the existing experimental data, which will require the frequency of the acoustical magnon, it is possible to estimate its boundaries.
No acoustical modes were observed above 10~\cm $\approx$ 1.24~meV according to~\cite{prosnikov_pbfe_2016}, thus it can be used to estimate a higher boundary of SIA.
The lower one could be estimated by taking into account the absence of acoustic mode up to 140 GHz $\approx$ 4.17~\cm $\approx$ 0.58~meV in antiferromagnetic resonance (AFMR) experiments~\cite{pankrats_pbfe_2014}.
Thus using the set of the exchange constants from~\cref{table:constants} and with~\cref{eq:sw_modes} will get anisotropy bounds of $-0.015 < \text{D} < -0.0033$ meV.
The validity of the obtained analytical results was confirmed by numerical calculations with SpinW library~\cite{sandor_toth_2017_838034,toth_linear_2015}.

\subsection{Two-magnon scattering}\label{sub:two_mag}

The most prominent feature observed in the magnetic Raman scattering spectra~\cite{prosnikov_pbfe_2016} below \TN is the broad and complex-shaped band attributed to the two-magnon scattering process due to its spectral and temperature-dependent characteristics.
First, we will start with the selection rules.
The effective two-magnon Raman Hamiltonian can be written according to the exchange scattering Fleury-Loudon mechanism~\cite{fleury_scattering_1968}:

\begin{equation} \label{eq:Fleury_Loudon_mechanism}
    \mathcal{H}_R \propto \sum_{i,d} (\mathbf{e}_{I} \cdot \mathbf{d})(\mathbf{e}_{S} \cdot \mathbf{d}) \mathbf{S}_i \mathbf{S}_j\ ,
\end{equation}

where $\boldsymbol{e}_{I}$ and $\boldsymbol{e}_{S}$ denote the polarization vectors of the incident and scattered light, and $\boldsymbol{d}$ is the vector connecting $i$-th ion with its nearest-neighbor for the specific exchange coupling.
Thus, taking into account only dominant intrachain J$_0$ interaction, this analysis predicts nonzero two-magnon scattering intensity only in ($bb$) polarizations, the case where both incident and scattered light polarized along the chains, which was, indeed, observed in the experiment~\cite{prosnikov_pbfe_2016}.

\begin{figure}
    \centering\includegraphics[width=8.6cm]{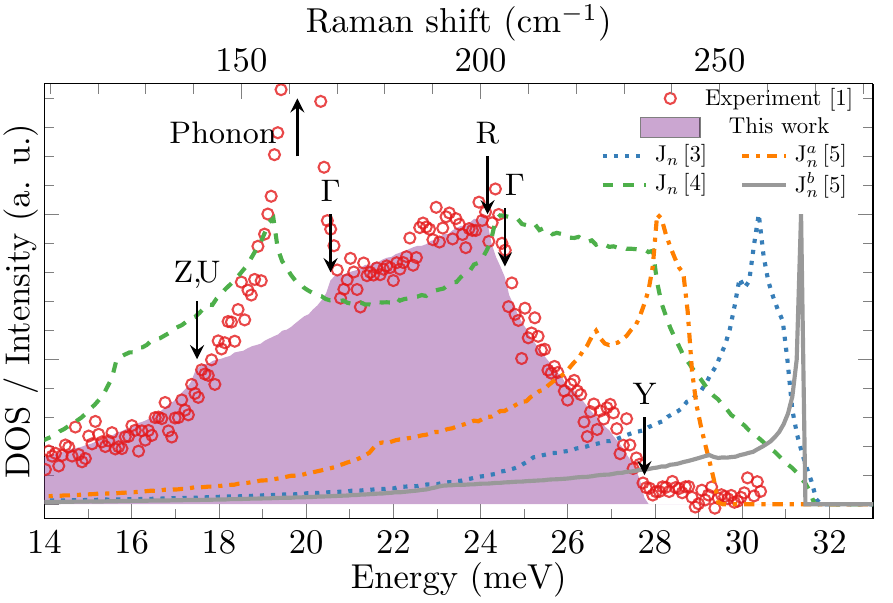}
    \caption{
        \label{fig:2M}
        Comparison of experimental Raman scattering spectra of two-magnon band (red marks, data extracted from Fig.~5 in~\cite{prosnikov_pbfe_2016}) with the spin-wave density of states calculated with exchange constants from~\cite{koo_density_2009,xiang_intra-chain_2016,curti_PbFe_2019}.
        Note that DOS energy scale is doubled to match 2M excitation.
        Arrows indicate Van Hove singularities in DOS with corresponding critical points in the Brillouin zone.
    }
\end{figure}

It is known that two-magnon excitations observed, e.g., by Raman scattering, reflect the spin-wave density of states (DOS)~\cite{fleury_scattering_1968}, which can be directly calculated based on dispersion relations~\cref{eq:sw_modes}.
It is necessary to use the full form of Hamiltonian~\cref{eq:PbFe_Hamiltonian} including all the exchange interactions to calculate the energy-depended shape of the two-magnon band.
Density of states is calculated according to:

\begin{equation} \label{eq:DOS}
    DOS=\oint\limits_{\mathclap{E(x,y,z)=\epsilon}} \frac{dS}{|\nabla E(x,y,z)|}\ ,
\end{equation}

where integral is taken numerically through constant energy surfaces within the first Brillouin zone.
The results of the calculations for different sets of exchange couplings from~\cref{table:constants} in comparison with the experiment are shown in~\cref{fig:2M}.
Note how drastically the shape is affected by $\text{J}_1$, and $\text{J}_2$, which allows us to undoubtedly determine them.

This approach, along with a proposed set of the optimized exchange constants $\text{J}_0=1.67, \text{J}_1=-0.18, \text{J}_2=0.094$~meV  allowed us to capture all essential experimental observations~\cite{prosnikov_pbfe_2016} such as
(i) high energy cut-off of the band at 28~meV
(ii) characteristic curvature of the band in the 24--28~meV range
(iii) nearly linear DOS of maximal energy in within 21--24~meV
(iv) nonzero and non-linear low-energy tail for energies less than 18~meV.

The limiting factors of further refinement of the exchange constants using 2M band are the presence of the intense phonon with the same A$_\text{g}$ symmetry with the energy of $\approx$19.5 meV, overshadowing part of the expected singularities, a rather noisy spectrum, and the absence of quantitative information on the lower energy tail.

It should be noted that one-magnon DOS surprisingly well describes the experimentally observed two-magnon band.
Usually, it is apparently different due to the magnon-magnon interactions~\cite{elliott_effects_1969} that strongly dampen and shift such excitations, which was demonstrated in paradigmatic examples of \ch{NiO}~\cite{dietz_infrared_1971} and \ch{RbMnF3}~\cite{fleury_evidence_1968}.
The negligible role of the magnon-magnon interactions could mean an increased lifetime of spin excitations~\cite{zhitomirsky_colloquium_2013}, which is a highly desirable goal for practical applications in antiferromagnetic spintronics~\cite{nemec_antiferromagnetic_2018}.
An interesting aspect is the presence of the pronounced Van Hove singularities in the 2M band, which can provide access to magnons in the specific points in the Brillouin zone and could be potentially applied to modulate exchange interactions and to control magnetic order in ultrafast timescales~\cite{mentink_manipulating_2017,batignani_probing_2015}.
Another striking feature is the accidental degeneracy of the high-energy (12.4~meV) optical magnon branch with A$_g$ phonon in the low-temperature limit, which could be used for spin dynamics manipulation through optical phonon pumping~\cite{nova_effective_2017}.

\subsection{Comparison with \abin calculations}\label{sub:abin}

Realization of both ferromagnetic and antiferromagnetic structures for different magnetic ions without change of the crystal symmetry in \Fam family spark the interest, and a few computational works~\cite{koo_density_2009,curti_PbFe_2019,xiang_intra-chain_2016} were done to shed light on this phenomenon.
Lattice dynamics was addressed in~\cite{murshed_anisotropic_2014,curti_PbFe_2019} and, in general, shows a good match with experimental data both on powdered samples~\cite{murshed_anisotropic_2014} and single crystals~\cite{prosnikov_pbfe_2016}.

Exchange constants up to the third neighbur were directly calculated in~\cite{koo_density_2009,curti_PbFe_2019} and, using energy mapping analysis (Eq.~3 in~\cite{koo_density_2009}), it is possible to extract constants from the energies of the magnetic structures from~\cite{xiang_intra-chain_2016}, which all summarized in~\cref{table:constants} and graphically shown in~\cref{fig:phase_diagram}.
All calculations agree on the dominant role of the intrachain exchange (J$_0$).
The J$_1$ is either antiferromagnetic or zero in the case of Gibbs free energy calculations~\cite{curti_PbFe_2019} and smaller than J$_2$, which is also antiferromagnetic.
It was shown that exchange constants are strongly dependent on Hubbard parameter U in DFT+U scheme~\cite{koo_density_2009}.
The comparison of the sets, with the experimental 2M band presented in~\cref{fig:2M}, clearly showing interchain coupling sensitivity, and substantial deviation for all the \abin sets.

Besides the direct determination of the exchange constants based on one- and two-magnon excitations, it is possible to use static magnetic data as an additional consistency check.
For example, Curie-Weiss temperature (sometimes referred to as Weiss or paramagnetic Curie temperature), which is the arithmetic average of all the exchange constants in the system~\cite{czachor_paramagnetic_1995}, could be used.
Taking into account the number and symmetry of the exchange couplings, it can be calculated as:

\begin{equation} \label{eq:Weiss_temperature}
    \Theta=-\frac{2}{3}\text{S}(\text{S}+1)[2 \text{J}_0 + 4 \text{J}_1 + 8 \text{J}_2]/k_\text{B}\ .
\end{equation}

This parameter could be experimentally determined from the temperature dependence of the magnetic susceptibility for $T >$~\TN.
This was done experimentally~\cite{pankrats_pbfe_2014}, and the reported values are negative, suggesting predominantly antiferromagnetic interactions in the system, and slightly anisotropic $\Theta_a=-256$, $\Theta_b=-272$, $\Theta_c=-262$~K.
Most of the sets, proposed in \abin works~\cite{koo_density_2009,curti_PbFe_2019,xiang_intra-chain_2016} overestimate $\Theta$ by 54--81~\%.
In contrast the optimal set of the exchange constants results in much closer value $\Theta = -228$~K.
The deviation of $\Theta$ from the proposed set could be explained by additional unaccounted superexchange couplings beyond J$_2$ or by the contribution of non-isotropic exchange interactions (see \cref{sub:beyond}).

The calculated exchange parameters in comparison with previously suggested ones are summarized in Table~\ref{table:constants}.
We hope that the proposed set of exchange constants, compatible with all up to date experimental observations, will be used for a systematic search of $U$ parameter in such a challenging system as \Fe.

\begin{table*}
    \centering
    \begin{tabular}{l|c|c|c|c|c|c|c}
    \hline
    \hline
        & J$_0$ & J$_1$ & J$_2$ & OM$_{calc}$ & OM$_{exp}$ & $\Theta_{calc}$ & $\Theta_{exp}$ \\
    \hline
    This work & 1.67 & $-0.18$ & 0.094 & 12.28 & -- & $-228$ & -- \\
    Koo \textit{et al.}~\cite{koo_density_2009} & 1.81 & 0.03447 & 0.3361 & 14.95 & -- & $-436$ & -- \\
    Xiang \textit{et al.}~\cite{xiang_intra-chain_2016} & 2.321 & 0.1815 & 0.20775 & 9.64 & -- & $-476$ & -- \\
    Curti \textit{et al.}~\cite{curti_PbFe_2019}$^2$ & 1.896 & 0.03447 & 0.259 & 13.35 & -- & $-406$ & -- \\
    Curti \textit{et al.}~\cite{curti_PbFe_2019}$^3$ & 2.1285 & 0.0 & 0.2499 & 14.64 & -- & $-423$ & -- \\
    Pankrats \textit{et al.}~\cite{pankrats_pbfe_2014} & -- & -- & -- & -- & -- & -- & $-263^1$ \\
    Park \textit{et al.}~\cite{park_synthesis_2003} & 2.24 & -- & -- & 11.2 & -- & $-303$ & -- \\
    Prosnikov \textit{et al.}~\cite{prosnikov_pbfe_2016} & 2.23 & -- & -- & 11.15 & 12.4 & -- & -- \\
    \hline
    \hline
    \multicolumn{8}{l}{$^{1}$\footnotesize{Averaged value, anisotropic ones are $\Theta_a = -256$~K, $\Theta_b = -272$~K, $\Theta_c = -262$~K~\cite{pankrats_pbfe_2014}}} \\
    \multicolumn{8}{l}{$^{2, 3}$\footnotesize{For only electronic energy and additional terms, respectively. For the details see Table~1 in~\cite{curti_PbFe_2019}.}} \\
    \end{tabular}
    \caption{Comparison of the exchange constants (meV, $\text{J}>0$ corresponds to AFM) and derived parameters, such as energies of the optical magnon branch (OM, meV, calculated according to~\cref{eq:sw_modes}), and Weiss temperatures $\Theta_{calc}$ (K, calculated according to~\cref{eq:Weiss_temperature}).
    }
    \label{table:constants}
\end{table*}

\subsection{Beyond isotropic exchange}\label{sub:beyond}

The unexplained anomaly was reported in~\cite{pankrats_pbfe_2014}, where unusual for typical easy-axis antiferromagnet peak-like maximum in the magnetic susceptibility for $H \parallel b$ geometry was measured.
Thus the potential presence of the anisotropic exchange couplings terms, such as Dzyaloshinskii-Moriya interaction, should be discussed.
In most cases, such interaction leads to a spin canting resulting in similar anomalies in susceptibility and a presence of the weak macroscopic magnetic moment in antiferromagnets~\cite{moriya_anisotropic_1960,moskvin_dzyaloshinskiimoriya_2019}, like in some well-known cases such as \ch{FeBO3}~\cite{dmitrienko_measuring_2014}, \ch{LiCoPO4}~\cite{fogh_dzyaloshinskii-moriya_2019}, perovskite manganites~\cite{sergienko_role_2006}, etc.
However, no magnetic moment was registered in \Fe for all the available field ranges and geometries~\cite{pankrats_pbfe_2014}.

It is known that DMI is governed by lattice symmetry~\cite{moriya_anisotropic_1960}.
Due to the presence of a mirror plane $(.m.)$ perpendicular to the J$_0$ exchange path and passing through its center, only $[\text{DM}_\text{x}, 0 ,\text{DM}_\text{z}]$ components of the DMI vector are allowed.
The DM$_\text{x}$ will lead to a slight canting of the spins within $bc$-planes inducing small ferromagnetic moment along individual chains.
However, due to the symmetry of the lattice, this canting is compensated by the same moment with the opposite direction from neighboring chains.
This canting is fully compatible with previously suggested magnetic space group $Pnma$~\cite{prosnikov_pbfe_2016} and can be described with $\Psi_x$ basis function.
On the other hand, the contribution of the DM$_\text{z}$ is negligible due to the orientation of the magnetic moments along the same axis.

Thus, the contribution of the antisymmetric exchange interaction with the DM$_\text{x}$ component can simultaneously explain the kink-like anomaly in magnetic susceptibility for $H\!\parallel\!b$ at \TN and the absence of weak ferromagnetic moment at lower temperatures.
This interaction will also directly affect spin dynamics in the form of magnon degeneracy lifting even without external magnetic field presence.
Numerical estimation of the acoustic mode splitting by the DMI with the semi-arbitrary value of 0.167~meV (1/10 of J$_0$) leads to 2.46~\cm $\approx$ 0.3~meV splitting of the acoustic mode which should be experimentally detectable with reasonably high-resolution Raman or IR spectroscopy setups.

\subsection{Magnetic structure dimensionality}\label{sub:dimension}

The assumption of the (quasi) one-dimensional (1D) magnetism in \Fe comes naturally considering its crystal structure, consistent from the well-separated chains of \ch{[FeO6]} octahedra running along the $b$-axis~\cite{park_synthesis_2003}.
The broad features above \TN in dc magnetic susceptibility on powdered samples~\cite{park_synthesis_2003} was also considered as a manifestation of short-range ordering characteristic for low-dimensional magnetic systems.
However, the detailed susceptibility investigation on the high-quality single crystals~\cite{pankrats_pbfe_2014} showed that broad features were caused by $\alpha$-\ch{Fe2O3} contamination, and $\chi$ behaves closer to three-dimensional Heisenberg antiferromagnet (except kink anomaly which was discussed in~\cref{sub:beyond}).

As the magnetic dimensionality measure, ratios of the intra- to interchain exchange couplings, taking into account coordination number ($z_n$), could be used~\cite{lemmens_magnetic_2003}.
The set of the optimal constants gives following values:
$1 : 0.22 : 0.22$ of $|\text{J}|_0*z_0 : |{J}_1|*z_1 : |{J}_2|*z_2$
which is closer to a 3D case, in comparison with other well-known one-dimensional systems as TTF-CuBDT~\cite{jeandey_TFF_1980}, \ch{CuGeO3}~\cite{nishi_CuGeO_1994}, and \ch{KCuF3}~\cite{hirakawa_KCuF_1970} with ratios $J_{intra}/J_{inter} << 1$.
Moreover, in most 1D systems, only single, nearest neighbour ($NN$) interchain coupling is considered important to capture spin dynamics properties, while for \Fe, all interaction up to third neighbour ($NN$, $NNN$, and $NNNN$) couplings are essential.

Thus, based on the above, \Fe should be considered a three-dimensional antiferromagnet in the low-temperature limit $T <<$~\TN. 
However, according to the magnetic susceptibility anomaly observed in~\cite{pankrats_pbfe_2014} and intense magnetic quasi-elastic scattering observed in polarization along the chains~\cite{prosnikov_pbfe_2016}, there is a possibility of quasi-one-dimensional behavior manifestation in the vicinity of the phase transition.

\subsection{Spin dynamics of \Cr}\label{sub:cr}
In comparison with \Fe, the magnetic structure of \Cr differs only in the direction of the easy axis~\cite{park_synthesis_2003}; thus, all previously derived equations and conclusions could be directly applied for an estimation of the energy of one- and two-magnon excitations.
Considering more than one order of magnitude lower \TN = 8\,K, and $\Theta = 45$\,K, the exchange constants are expected to be proportionally smaller.
However, up to now, there are no published data on the spin dynamics of \Cr.

According to~\cite{koo_density_2009} the exchange parameters of the optimized structure (calculated with Hubbard parameter $U = 2.0$\,eV) are $\text{J}_0 \approx 0.52$\,meV, $\text{J}_1 \approx -0.0345$\,meV, and $\text{J}_2 \approx 0.069$\,meV, which with reduced, in comparison with \Fe, the spin value of \ch{Cr^3+} ions $S = 3/2$ gives expected energy of two-magnon excitation band maximum of $\approx 5\,\text{meV}$ and the energy of optical magnon mode at $\approx 2.7~\text{meV}$ which are both accessible in typical low-energy Raman scattering experiments.

\section{Summary and conclusion}\label{sec:concl}
With the use of linear spin-wave theory, the closed-form of the magnon dispersion relation of \Fe was derived, including exchange couplings up to the third neighbor and single-ion anisotropy of the easy axis type.
It is demonstrated that magnetic excitations observed in Raman scattering~\cite{prosnikov_pbfe_2016} are optical (exchange) magnon and two-magnon band and based on their energy and magnetic susceptibility data~\cite{pankrats_pbfe_2014}, the consistent set of exchange coupling constants is proposed: $\text{J}_0=1.67, \text{J}_1=-0.18, \text{J}_2=0.094$~meV.
It is shown that \abin calculations~\cite{koo_density_2009,curti_PbFe_2019,xiang_intra-chain_2016} overestimate both J$_0$ and J$_2$ while predicts the opposite sign for J$_1$.
Nonzero components of the Dzyaloshinskii-Moriya interaction are allowed for the J$_0$ exchange path, which could be responsible for the magnetic susceptibility anomaly~\cite{pankrats_pbfe_2014}.
Surprisingly, the shape of the two-magnon band is well described by the one-magnon density of states, which indicates a vanishingly small role of the magnon-magnon interactions.
We hope that obtained results will stimulate both experimental research, such as IR and low-energy Raman spectroscopy to find acoustic modes and inelastic neutron scattering to directly probe magnon dispersion, and theoretical ones on spin dynamics and a systematic study of the exchange coupling constants dependence of Hubbard parameter ($U$) for \abin calculations.

\begin{acknowledgments}
The support of A.\,M~Kalashnikova, fruitful discussions with Beatrice~T.~Crow, and assistance with the carbon tube by M.~Berben are greatly acknowledged.
We also acknowledge the support of the HFML-RU/NWO-I, member of the European Magnetic Field Laboratory (EMFL).
\end{acknowledgments}



\bibliography{PbFe_PRB.bib}

\begin{thebibliography}{40}%
\makeatletter
\providecommand \@ifxundefined [1]{%
 \@ifx{#1\undefined}
}%
\providecommand \@ifnum [1]{%
 \ifnum #1\expandafter \@firstoftwo
 \else \expandafter \@secondoftwo
 \fi
}%
\providecommand \@ifx [1]{%
 \ifx #1\expandafter \@firstoftwo
 \else \expandafter \@secondoftwo
 \fi
}%
\providecommand \natexlab [1]{#1}%
\providecommand \enquote  [1]{``#1''}%
\providecommand \bibnamefont  [1]{#1}%
\providecommand \bibfnamefont [1]{#1}%
\providecommand \citenamefont [1]{#1}%
\providecommand \href@noop [0]{\@secondoftwo}%
\providecommand \href [0]{\begingroup \@sanitize@url \@href}%
\providecommand \@href[1]{\@@startlink{#1}\@@href}%
\providecommand \@@href[1]{\endgroup#1\@@endlink}%
\providecommand \@sanitize@url [0]{\catcode `\\12\catcode `\$12\catcode
  `\&12\catcode `\#12\catcode `\^12\catcode `\_12\catcode `\%12\relax}%
\providecommand \@@startlink[1]{}%
\providecommand \@@endlink[0]{}%
\providecommand \url  [0]{\begingroup\@sanitize@url \@url }%
\providecommand \@url [1]{\endgroup\@href {#1}{\urlprefix }}%
\providecommand \urlprefix  [0]{URL }%
\providecommand \Eprint [0]{\href }%
\providecommand \doibase [0]{https://doi.org/}%
\providecommand \selectlanguage [0]{\@gobble}%
\providecommand \bibinfo  [0]{\@secondoftwo}%
\providecommand \bibfield  [0]{\@secondoftwo}%
\providecommand \translation [1]{[#1]}%
\providecommand \BibitemOpen [0]{}%
\providecommand \bibitemStop [0]{}%
\providecommand \bibitemNoStop [0]{.\EOS\space}%
\providecommand \EOS [0]{\spacefactor3000\relax}%
\providecommand \BibitemShut  [1]{\csname bibitem#1\endcsname}%
\let\auto@bib@innerbib\@empty
\bibitem [{\citenamefont {Prosnikov}\ \emph {et~al.}(2016)\citenamefont
  {Prosnikov}, \citenamefont {Smirnov}, \citenamefont {Davydov}, \citenamefont
  {Sablina},\ and\ \citenamefont {Pisarev}}]{prosnikov_pbfe_2016}%
  \BibitemOpen
  \bibfield  {author} {\bibinfo {author} {\bibfnamefont {M.~A.}\ \bibnamefont
  {Prosnikov}}, \bibinfo {author} {\bibfnamefont {A.~N.}\ \bibnamefont
  {Smirnov}}, \bibinfo {author} {\bibfnamefont {V.~Y.}\ \bibnamefont
  {Davydov}}, \bibinfo {author} {\bibfnamefont {K.~A.}\ \bibnamefont
  {Sablina}},\ and\ \bibinfo {author} {\bibfnamefont {R.~V.}\ \bibnamefont
  {Pisarev}},\ }\bibfield  {title} {\bibinfo {title} {Lattice and magnetic
  dynamics of a quasi-one-dimensional chain antiferromagnet {PbFeBO$_4$}},\
  }\href {https://doi.org/10.1088/0953-8984/29/2/025808} {\bibfield  {journal}
  {\bibinfo  {journal} {Journal of Physics: Condensed Matter}\ }\textbf
  {\bibinfo {volume} {29}},\ \bibinfo {pages} {025808} (\bibinfo {year}
  {2016})}\BibitemShut {NoStop}%
\bibitem [{\citenamefont {Pankrats}\ \emph {et~al.}(2014)\citenamefont
  {Pankrats}, \citenamefont {Sablina}, \citenamefont {Velikanov}, \citenamefont
  {Vorotynov}, \citenamefont {Bayukov}, \citenamefont {Eremin}, \citenamefont
  {Molokeev}, \citenamefont {Popkov},\ and\ \citenamefont
  {Krasikov}}]{pankrats_pbfe_2014}%
  \BibitemOpen
  \bibfield  {author} {\bibinfo {author} {\bibfnamefont {A.}~\bibnamefont
  {Pankrats}}, \bibinfo {author} {\bibfnamefont {K.}~\bibnamefont {Sablina}},
  \bibinfo {author} {\bibfnamefont {D.}~\bibnamefont {Velikanov}}, \bibinfo
  {author} {\bibfnamefont {A.}~\bibnamefont {Vorotynov}}, \bibinfo {author}
  {\bibfnamefont {O.}~\bibnamefont {Bayukov}}, \bibinfo {author} {\bibfnamefont
  {A.}~\bibnamefont {Eremin}}, \bibinfo {author} {\bibfnamefont
  {M.}~\bibnamefont {Molokeev}}, \bibinfo {author} {\bibfnamefont
  {S.}~\bibnamefont {Popkov}},\ and\ \bibinfo {author} {\bibfnamefont
  {A.}~\bibnamefont {Krasikov}},\ }\bibfield  {title} {\bibinfo {title}
  {Magnetic and dielectric properties of the {PbFeBO$_4$} single crystal},\
  }\href {https://doi.org/10.1016/j.jmmm.2013.10.018} {\bibfield  {journal}
  {\bibinfo  {journal} {Journal of Magnetism and Magnetic Materials}\ }\textbf
  {\bibinfo {volume} {353}},\ \bibinfo {pages} {23} (\bibinfo {year}
  {2014})}\BibitemShut {NoStop}%
\bibitem [{\citenamefont {Koo}\ and\ \citenamefont
  {Whangbo}(2009)}]{koo_density_2009}%
  \BibitemOpen
  \bibfield  {author} {\bibinfo {author} {\bibfnamefont {H.-J.}\ \bibnamefont
  {Koo}}\ and\ \bibinfo {author} {\bibfnamefont {M.-H.}\ \bibnamefont
  {Whangbo}},\ }\bibfield  {title} {\bibinfo {title} {Density functional
  investigation of the magnetic properties of {PbMBO$_4$} ({M}= {Cr}, {Mn},
  {Fe})},\ }\href {https://doi.org/10.1016/j.ssc.2009.01.030} {\bibfield
  {journal} {\bibinfo  {journal} {Solid State Communications}\ }\textbf
  {\bibinfo {volume} {149}},\ \bibinfo {pages} {602} (\bibinfo {year}
  {2009})}\BibitemShut {NoStop}%
\bibitem [{\citenamefont {Xiang}\ \emph {et~al.}(2016)\citenamefont {Xiang},
  \citenamefont {Tang}, \citenamefont {Zhang},\ and\ \citenamefont
  {He}}]{xiang_intra-chain_2016}%
  \BibitemOpen
  \bibfield  {author} {\bibinfo {author} {\bibfnamefont {H.}~\bibnamefont
  {Xiang}}, \bibinfo {author} {\bibfnamefont {Y.}~\bibnamefont {Tang}},
  \bibinfo {author} {\bibfnamefont {S.}~\bibnamefont {Zhang}},\ and\ \bibinfo
  {author} {\bibfnamefont {Z.}~\bibnamefont {He}},\ }\bibfield  {title}
  {\bibinfo {title} {Intra-chain superexchange couplings in quasi-{1D} 3$d$
  transition-metal magnetic compounds},\ }\href
  {https://doi.org/10.1088/0953-8984/28/27/276003} {\bibfield  {journal}
  {\bibinfo  {journal} {Journal of Physics: Condensed Matter}\ }\textbf
  {\bibinfo {volume} {28}},\ \bibinfo {pages} {276003} (\bibinfo {year}
  {2016})}\BibitemShut {NoStop}%
\bibitem [{\citenamefont {Curti}\ \emph
  {et~al.}(2019{\natexlab{a}})\citenamefont {Curti}, \citenamefont {Murshed},
  \citenamefont {Bredow}, \citenamefont {Bahnemann}, \citenamefont {Gesing},\
  and\ \citenamefont {Mendive}}]{curti_PbFe_2019}%
  \BibitemOpen
  \bibfield  {author} {\bibinfo {author} {\bibfnamefont {M.}~\bibnamefont
  {Curti}}, \bibinfo {author} {\bibfnamefont {M.~M.}\ \bibnamefont {Murshed}},
  \bibinfo {author} {\bibfnamefont {T.}~\bibnamefont {Bredow}}, \bibinfo
  {author} {\bibfnamefont {D.~W.}\ \bibnamefont {Bahnemann}}, \bibinfo {author}
  {\bibfnamefont {T.~M.}\ \bibnamefont {Gesing}},\ and\ \bibinfo {author}
  {\bibfnamefont {C.~B.}\ \bibnamefont {Mendive}},\ }\bibfield  {title}
  {\bibinfo {title} {Elastic, phononic, magnetic and electronic properties of
  quasi-one-dimensional {PbFeBO$_4$}},\ }\bibfield  {journal} {\bibinfo
  {journal} {Journal of Materials Science}\ }\href
  {https://doi.org/10.1007/s10853-019-03866-1} {10.1007/s10853-019-03866-1}
  (\bibinfo {year} {2019}{\natexlab{a}})\BibitemShut {NoStop}%
\bibitem [{\citenamefont {Němec}\ \emph {et~al.}(2018)\citenamefont {Němec},
  \citenamefont {Fiebig}, \citenamefont {Kampfrath},\ and\ \citenamefont
  {Kimel}}]{nemec_antiferromagnetic_2018}%
  \BibitemOpen
  \bibfield  {author} {\bibinfo {author} {\bibfnamefont {P.}~\bibnamefont
  {Němec}}, \bibinfo {author} {\bibfnamefont {M.}~\bibnamefont {Fiebig}},
  \bibinfo {author} {\bibfnamefont {T.}~\bibnamefont {Kampfrath}},\ and\
  \bibinfo {author} {\bibfnamefont {A.~V.}\ \bibnamefont {Kimel}},\ }\bibfield
  {title} {\bibinfo {title} {Antiferromagnetic opto-spintronics},\ }\href
  {https://doi.org/10.1038/s41567-018-0051-x} {\bibfield  {journal} {\bibinfo
  {journal} {Nature Physics}\ }\textbf {\bibinfo {volume} {14}},\ \bibinfo
  {pages} {229} (\bibinfo {year} {2018})}\BibitemShut {NoStop}%
\bibitem [{\citenamefont {Gomonay}\ and\ \citenamefont
  {Loktev}(2014)}]{gomonay_spintronics_2014}%
  \BibitemOpen
  \bibfield  {author} {\bibinfo {author} {\bibfnamefont {E.~V.}\ \bibnamefont
  {Gomonay}}\ and\ \bibinfo {author} {\bibfnamefont {V.~M.}\ \bibnamefont
  {Loktev}},\ }\bibfield  {title} {\bibinfo {title} {Spintronics of
  antiferromagnetic systems},\ }\href {https://doi.org/10.1063/1.4862467}
  {\bibfield  {journal} {\bibinfo  {journal} {Low Temperature Physics}\
  }\textbf {\bibinfo {volume} {40}},\ \bibinfo {pages} {17} (\bibinfo {year}
  {2014})}\BibitemShut {NoStop}%
\bibitem [{\citenamefont {Baltz}\ \emph {et~al.}(2018)\citenamefont {Baltz},
  \citenamefont {Manchon}, \citenamefont {Tsoi}, \citenamefont {Moriyama},
  \citenamefont {Ono},\ and\ \citenamefont
  {Tserkovnyak}}]{baltz_antiferromagnetic_2018}%
  \BibitemOpen
  \bibfield  {author} {\bibinfo {author} {\bibfnamefont {V.}~\bibnamefont
  {Baltz}}, \bibinfo {author} {\bibfnamefont {A.}~\bibnamefont {Manchon}},
  \bibinfo {author} {\bibfnamefont {M.}~\bibnamefont {Tsoi}}, \bibinfo {author}
  {\bibfnamefont {T.}~\bibnamefont {Moriyama}}, \bibinfo {author}
  {\bibfnamefont {T.}~\bibnamefont {Ono}},\ and\ \bibinfo {author}
  {\bibfnamefont {Y.}~\bibnamefont {Tserkovnyak}},\ }\bibfield  {title}
  {\bibinfo {title} {Antiferromagnetic spintronics},\ }\href
  {https://doi.org/10.1103/revmodphys.90.015005} {\bibfield  {journal}
  {\bibinfo  {journal} {Reviews of Modern Physics}\ }\textbf {\bibinfo {volume}
  {90}},\ \bibinfo {pages} {015005} (\bibinfo {year} {2018})}\BibitemShut
  {NoStop}%
\bibitem [{\citenamefont {Jungwirth}\ \emph {et~al.}(2016)\citenamefont
  {Jungwirth}, \citenamefont {Marti}, \citenamefont {Wadley},\ and\
  \citenamefont {Wunderlich}}]{jungwirth_antiferromagnetic_2016}%
  \BibitemOpen
  \bibfield  {author} {\bibinfo {author} {\bibfnamefont {T.}~\bibnamefont
  {Jungwirth}}, \bibinfo {author} {\bibfnamefont {X.}~\bibnamefont {Marti}},
  \bibinfo {author} {\bibfnamefont {P.}~\bibnamefont {Wadley}},\ and\ \bibinfo
  {author} {\bibfnamefont {J.}~\bibnamefont {Wunderlich}},\ }\bibfield  {title}
  {\bibinfo {title} {Antiferromagnetic spintronics},\ }\href
  {https://doi.org/10.1038/nnano.2016.18} {\bibfield  {journal} {\bibinfo
  {journal} {Nature Nanotechnology}\ }\textbf {\bibinfo {volume} {11}},\
  \bibinfo {pages} {231} (\bibinfo {year} {2016})}\BibitemShut {NoStop}%
\bibitem [{\citenamefont {Spaldin}\ and\ \citenamefont
  {Ramesh}(2019)}]{spaldin_advances_2019}%
  \BibitemOpen
  \bibfield  {author} {\bibinfo {author} {\bibfnamefont {N.~A.}\ \bibnamefont
  {Spaldin}}\ and\ \bibinfo {author} {\bibfnamefont {R.}~\bibnamefont
  {Ramesh}},\ }\bibfield  {title} {\bibinfo {title} {Advances in
  magnetoelectric multiferroics},\ }\href
  {https://doi.org/10.1038/s41563-018-0275-2} {\bibfield  {journal} {\bibinfo
  {journal} {Nature Materials}\ }\textbf {\bibinfo {volume} {18}},\ \bibinfo
  {pages} {203} (\bibinfo {year} {2019})}\BibitemShut {NoStop}%
\bibitem [{\citenamefont {Son}\ \emph {et~al.}(2019)\citenamefont {Son},
  \citenamefont {Park}, \citenamefont {Kim}, \citenamefont {Cho}, \citenamefont
  {Kim}, \citenamefont {Sandilands}, \citenamefont {Sohn}, \citenamefont
  {Park}, \citenamefont {Moon},\ and\ \citenamefont
  {Noh}}]{son_unconventional_2019}%
  \BibitemOpen
  \bibfield  {author} {\bibinfo {author} {\bibfnamefont {J.}~\bibnamefont
  {Son}}, \bibinfo {author} {\bibfnamefont {B.~C.}\ \bibnamefont {Park}},
  \bibinfo {author} {\bibfnamefont {C.~H.}\ \bibnamefont {Kim}}, \bibinfo
  {author} {\bibfnamefont {H.}~\bibnamefont {Cho}}, \bibinfo {author}
  {\bibfnamefont {S.~Y.}\ \bibnamefont {Kim}}, \bibinfo {author} {\bibfnamefont
  {L.~J.}\ \bibnamefont {Sandilands}}, \bibinfo {author} {\bibfnamefont
  {C.}~\bibnamefont {Sohn}}, \bibinfo {author} {\bibfnamefont {J.-G.}\
  \bibnamefont {Park}}, \bibinfo {author} {\bibfnamefont {S.~J.}\ \bibnamefont
  {Moon}},\ and\ \bibinfo {author} {\bibfnamefont {T.~W.}\ \bibnamefont
  {Noh}},\ }\bibfield  {title} {\bibinfo {title} {Unconventional spin-phonon
  coupling via the {Dzyaloshinskii}–{Moriya} interaction},\ }\href
  {https://doi.org/10.1038/s41535-019-0157-0} {\bibfield  {journal} {\bibinfo
  {journal} {npj Quantum Materials}\ }\textbf {\bibinfo {volume} {4}},\
  \bibinfo {pages} {1} (\bibinfo {year} {2019})}\BibitemShut {NoStop}%
\bibitem [{\citenamefont {Han}\ \emph {et~al.}(2019)\citenamefont {Han},
  \citenamefont {Lee},\ and\ \citenamefont {Moon}}]{han_lattice_2019}%
  \BibitemOpen
  \bibfield  {author} {\bibinfo {author} {\bibfnamefont {S.}~\bibnamefont
  {Han}}, \bibinfo {author} {\bibfnamefont {J.}~\bibnamefont {Lee}},\ and\
  \bibinfo {author} {\bibfnamefont {E.-G.}\ \bibnamefont {Moon}},\ }\bibfield
  {title} {\bibinfo {title} {Lattice vibration as a knob for novel quantum
  criticality: {Emergence} of supersymmetry from spin-lattice coupling},\
  }\href@noop {} {\bibfield  {journal} {\bibinfo  {journal} {arXiv:1911.01435
  [cond-mat, physics:hep-th]}\ } (\bibinfo {year} {2019})},\ \bibinfo {note}
  {arXiv: 1911.01435}\BibitemShut {NoStop}%
\bibitem [{\citenamefont {Fischer}\ and\ \citenamefont
  {Schneider}(2008)}]{fischer_crystal_2008}%
  \BibitemOpen
  \bibfield  {author} {\bibinfo {author} {\bibfnamefont {R.~X.}\ \bibnamefont
  {Fischer}}\ and\ \bibinfo {author} {\bibfnamefont {H.}~\bibnamefont
  {Schneider}},\ }\bibfield  {title} {\bibinfo {title} {Crystal chemistry of
  borates and borosilicates with mullite-type structures: a review},\ }\href
  {https://doi.org/10.1127/0935-1221/2008/0020-1831} {\bibfield  {journal}
  {\bibinfo  {journal} {European Journal of Mineralogy}\ }\textbf {\bibinfo
  {volume} {20}},\ \bibinfo {pages} {917} (\bibinfo {year} {2008})}\BibitemShut
  {NoStop}%
\bibitem [{\citenamefont {Murshed}\ \emph {et~al.}(2012)\citenamefont
  {Murshed}, \citenamefont {Fischer},\ and\ \citenamefont
  {Gesing}}]{murshed_role_2012}%
  \BibitemOpen
  \bibfield  {author} {\bibinfo {author} {\bibfnamefont {M.~M.}\ \bibnamefont
  {Murshed}}, \bibinfo {author} {\bibfnamefont {R.~X.}\ \bibnamefont
  {Fischer}},\ and\ \bibinfo {author} {\bibfnamefont {T.~M.}\ \bibnamefont
  {Gesing}},\ }\bibfield  {title} {\bibinfo {title} {The role of the
  {Pb$^{2+}$} lone electron pair for bond valence sum analysis in mullite-type
  {PbMBO$_4$} ({M} = {Al}, {Mn} and {Fe}) compounds},\ }\href
  {https://doi.org/10.1524/zkri.2012.1483} {\bibfield  {journal} {\bibinfo
  {journal} {Zeitschrift für Kristallographie - Crystalline Materials}\
  }\textbf {\bibinfo {volume} {227}},\ \bibinfo {pages} {580} (\bibinfo {year}
  {2012})}\BibitemShut {NoStop}%
\bibitem [{\citenamefont {Park}\ \emph {et~al.}(2003)\citenamefont {Park},
  \citenamefont {Lam}, \citenamefont {Greedan},\ and\ \citenamefont
  {Barbier}}]{park_synthesis_2003}%
  \BibitemOpen
  \bibfield  {author} {\bibinfo {author} {\bibfnamefont {H.}~\bibnamefont
  {Park}}, \bibinfo {author} {\bibfnamefont {R.}~\bibnamefont {Lam}}, \bibinfo
  {author} {\bibfnamefont {J.~E.}\ \bibnamefont {Greedan}},\ and\ \bibinfo
  {author} {\bibfnamefont {J.}~\bibnamefont {Barbier}},\ }\bibfield  {title}
  {\bibinfo {title} {Synthesis, {Crystal} {Structure}, {Crystal} {Chemistry},
  and {Magnetic} {Properties} of {PbMBO$_4$} ({M} = {Cr}, {Mn}, {Fe}): {A}
  {New} {Structure} {Type} {Exhibiting} {One}-{Dimensional} {Magnetism}},\
  }\href {https://doi.org/10.1021/cm0217452} {\bibfield  {journal} {\bibinfo
  {journal} {Chemistry of Materials}\ }\textbf {\bibinfo {volume} {15}},\
  \bibinfo {pages} {1703} (\bibinfo {year} {2003})}\BibitemShut {NoStop}%
\bibitem [{\citenamefont {Pankrats}\ \emph {et~al.}(2016)\citenamefont
  {Pankrats}, \citenamefont {Sablina}, \citenamefont {Eremin}, \citenamefont
  {Balaev}, \citenamefont {Kolkov}, \citenamefont {Tugarinov},\ and\
  \citenamefont {Bovina}}]{pankrats_pbmn_2016}%
  \BibitemOpen
  \bibfield  {author} {\bibinfo {author} {\bibfnamefont {A.}~\bibnamefont
  {Pankrats}}, \bibinfo {author} {\bibfnamefont {K.}~\bibnamefont {Sablina}},
  \bibinfo {author} {\bibfnamefont {M.}~\bibnamefont {Eremin}}, \bibinfo
  {author} {\bibfnamefont {A.}~\bibnamefont {Balaev}}, \bibinfo {author}
  {\bibfnamefont {M.}~\bibnamefont {Kolkov}}, \bibinfo {author} {\bibfnamefont
  {V.}~\bibnamefont {Tugarinov}},\ and\ \bibinfo {author} {\bibfnamefont
  {A.}~\bibnamefont {Bovina}},\ }\bibfield  {title} {\bibinfo {title}
  {Ferromagnetism and strong magnetic anisotropy of the {PbMnBO$_4$}
  orthoborate single crystals},\ }\href
  {https://doi.org/10.1016/j.jmmm.2016.04.042} {\bibfield  {journal} {\bibinfo
  {journal} {Journal of Magnetism and Magnetic Materials}\ }\textbf {\bibinfo
  {volume} {414}},\ \bibinfo {pages} {82} (\bibinfo {year} {2016})}\BibitemShut
  {NoStop}%
\bibitem [{\citenamefont {Murshed}\ \emph {et~al.}(2014)\citenamefont
  {Murshed}, \citenamefont {Mendive}, \citenamefont {Curti}, \citenamefont
  {Nénert}, \citenamefont {Kalita}, \citenamefont {Lipinska}, \citenamefont
  {Cornelius}, \citenamefont {Huq},\ and\ \citenamefont
  {Gesing}}]{murshed_anisotropic_2014}%
  \BibitemOpen
  \bibfield  {author} {\bibinfo {author} {\bibfnamefont {M.~M.}\ \bibnamefont
  {Murshed}}, \bibinfo {author} {\bibfnamefont {C.~B.}\ \bibnamefont
  {Mendive}}, \bibinfo {author} {\bibfnamefont {M.}~\bibnamefont {Curti}},
  \bibinfo {author} {\bibfnamefont {G.}~\bibnamefont {Nénert}}, \bibinfo
  {author} {\bibfnamefont {P.~E.}\ \bibnamefont {Kalita}}, \bibinfo {author}
  {\bibfnamefont {K.}~\bibnamefont {Lipinska}}, \bibinfo {author}
  {\bibfnamefont {A.~L.}\ \bibnamefont {Cornelius}}, \bibinfo {author}
  {\bibfnamefont {A.}~\bibnamefont {Huq}},\ and\ \bibinfo {author}
  {\bibfnamefont {T.~M.}\ \bibnamefont {Gesing}},\ }\bibfield  {title}
  {\bibinfo {title} {Anisotropic lattice thermal expansion of {PbFeBO4}: {A}
  study by {X}-ray and neutron diffraction, {Raman} spectroscopy and {DFT}
  calculations},\ }\href {https://doi.org/10.1016/j.materresbull.2014.07.005}
  {\bibfield  {journal} {\bibinfo  {journal} {Materials Research Bulletin}\
  }\textbf {\bibinfo {volume} {59}},\ \bibinfo {pages} {170} (\bibinfo {year}
  {2014})}\BibitemShut {NoStop}%
\bibitem [{\citenamefont {Curti}\ \emph
  {et~al.}(2019{\natexlab{b}})\citenamefont {Curti}, \citenamefont {Mendive},
  \citenamefont {Bredow}, \citenamefont {Murshed},\ and\ \citenamefont
  {Gesing}}]{curti_SnM_2019}%
  \BibitemOpen
  \bibfield  {author} {\bibinfo {author} {\bibfnamefont {M.}~\bibnamefont
  {Curti}}, \bibinfo {author} {\bibfnamefont {C.~B.}\ \bibnamefont {Mendive}},
  \bibinfo {author} {\bibfnamefont {T.}~\bibnamefont {Bredow}}, \bibinfo
  {author} {\bibfnamefont {M.~M.}\ \bibnamefont {Murshed}},\ and\ \bibinfo
  {author} {\bibfnamefont {T.~M.}\ \bibnamefont {Gesing}},\ }\bibfield  {title}
  {\bibinfo {title} {Structural, vibrational and electronic properties of
  {SnMBO$_4$} ({M = Al, Ga}): a predictive hybrid {DFT} study},\ }\href
  {https://doi.org/10.1088/1361-648X/ab20a1} {\bibfield  {journal} {\bibinfo
  {journal} {Journal of Physics: Condensed Matter}\ }\textbf {\bibinfo {volume}
  {31}},\ \bibinfo {pages} {345701} (\bibinfo {year}
  {2019}{\natexlab{b}})}\BibitemShut {NoStop}%
\bibitem [{\citenamefont {Tóth}(2017)}]{sandor_toth_2017_838034}%
  \BibitemOpen
  \bibfield  {author} {\bibinfo {author} {\bibfnamefont {S.}~\bibnamefont
  {Tóth}},\ }\href {https://doi.org/10.5281/zenodo.838034} {\bibinfo {title}
  {tsdev/spinw: pyspinw 3.0}} (\bibinfo {year} {2017})\BibitemShut {NoStop}%
\bibitem [{\citenamefont {Holstein}\ and\ \citenamefont
  {Primakoff}(1940)}]{holstein_field_1940}%
  \BibitemOpen
  \bibfield  {author} {\bibinfo {author} {\bibfnamefont {T.}~\bibnamefont
  {Holstein}}\ and\ \bibinfo {author} {\bibfnamefont {H.}~\bibnamefont
  {Primakoff}},\ }\bibfield  {title} {\bibinfo {title} {Field {Dependence} of
  the {Intrinsic} {Domain} {Magnetization} of a {Ferromagnet}},\ }\href
  {https://doi.org/10.1103/PhysRev.58.1098} {\bibfield  {journal} {\bibinfo
  {journal} {Physical Review}\ }\textbf {\bibinfo {volume} {58}},\ \bibinfo
  {pages} {1098} (\bibinfo {year} {1940})}\BibitemShut {NoStop}%
\bibitem [{\citenamefont {White}\ \emph {et~al.}(1965)\citenamefont {White},
  \citenamefont {Sparks},\ and\ \citenamefont
  {Ortenburger}}]{white_diagonalization_1965}%
  \BibitemOpen
  \bibfield  {author} {\bibinfo {author} {\bibfnamefont {R.~M.}\ \bibnamefont
  {White}}, \bibinfo {author} {\bibfnamefont {M.}~\bibnamefont {Sparks}},\ and\
  \bibinfo {author} {\bibfnamefont {I.}~\bibnamefont {Ortenburger}},\
  }\bibfield  {title} {\bibinfo {title} {Diagonalization of the
  {Antiferromagnetic} {Magnon}-{Phonon} {Interaction}},\ }\href
  {https://doi.org/10.1103/PhysRev.139.A450} {\bibfield  {journal} {\bibinfo
  {journal} {Physical Review}\ }\textbf {\bibinfo {volume} {139}},\ \bibinfo
  {pages} {A450} (\bibinfo {year} {1965})}\BibitemShut {NoStop}%
\bibitem [{\citenamefont {Toth}\ and\ \citenamefont
  {Lake}(2015)}]{toth_linear_2015}%
  \BibitemOpen
  \bibfield  {author} {\bibinfo {author} {\bibfnamefont {S.}~\bibnamefont
  {Toth}}\ and\ \bibinfo {author} {\bibfnamefont {B.}~\bibnamefont {Lake}},\
  }\bibfield  {title} {\bibinfo {title} {Linear spin wave theory for single-{Q}
  incommensurate magnetic structures},\ }\href
  {https://doi.org/10.1088/0953-8984/27/16/166002} {\bibfield  {journal}
  {\bibinfo  {journal} {Journal of Physics: Condensed Matter}\ }\textbf
  {\bibinfo {volume} {27}},\ \bibinfo {pages} {166002} (\bibinfo {year}
  {2015})}\BibitemShut {NoStop}%
\bibitem [{\citenamefont {Fleury}\ and\ \citenamefont
  {Loudon}(1968)}]{fleury_scattering_1968}%
  \BibitemOpen
  \bibfield  {author} {\bibinfo {author} {\bibfnamefont {P.~A.}\ \bibnamefont
  {Fleury}}\ and\ \bibinfo {author} {\bibfnamefont {R.}~\bibnamefont
  {Loudon}},\ }\bibfield  {title} {\bibinfo {title} {Scattering of {Light} by
  {One}- and {Two}-{Magnon} {Excitations}},\ }\href
  {https://doi.org/10.1103/PhysRev.166.514} {\bibfield  {journal} {\bibinfo
  {journal} {Physical Review}\ }\textbf {\bibinfo {volume} {166}},\ \bibinfo
  {pages} {514} (\bibinfo {year} {1968})}\BibitemShut {NoStop}%
\bibitem [{\citenamefont {Elliott}\ and\ \citenamefont
  {Thorpe}(1969)}]{elliott_effects_1969}%
  \BibitemOpen
  \bibfield  {author} {\bibinfo {author} {\bibfnamefont {R.~J.}\ \bibnamefont
  {Elliott}}\ and\ \bibinfo {author} {\bibfnamefont {M.~F.}\ \bibnamefont
  {Thorpe}},\ }\bibfield  {title} {\bibinfo {title} {The effects of
  magnon-magnon interaction on the two-magnon spectra of antiferromagnets},\
  }\href {https://doi.org/10.1088/0022-3719/2/9/312} {\bibfield  {journal}
  {\bibinfo  {journal} {Journal of Physics C: Solid State Physics}\ }\textbf
  {\bibinfo {volume} {2}},\ \bibinfo {pages} {1630} (\bibinfo {year}
  {1969})}\BibitemShut {NoStop}%
\bibitem [{\citenamefont {Dietz}\ \emph {et~al.}(1971)\citenamefont {Dietz},
  \citenamefont {Parisot},\ and\ \citenamefont
  {Meixner}}]{dietz_infrared_1971}%
  \BibitemOpen
  \bibfield  {author} {\bibinfo {author} {\bibfnamefont {R.~E.}\ \bibnamefont
  {Dietz}}, \bibinfo {author} {\bibfnamefont {G.~I.}\ \bibnamefont {Parisot}},\
  and\ \bibinfo {author} {\bibfnamefont {A.~E.}\ \bibnamefont {Meixner}},\
  }\bibfield  {title} {\bibinfo {title} {Infrared {Absorption} and {Raman}
  {Scattering} by {Two}-{Magnon} {Processes} in {NiO}},\ }\href
  {https://doi.org/10.1103/PhysRevB.4.2302} {\bibfield  {journal} {\bibinfo
  {journal} {Physical Review B}\ }\textbf {\bibinfo {volume} {4}},\ \bibinfo
  {pages} {2302} (\bibinfo {year} {1971})}\BibitemShut {NoStop}%
\bibitem [{\citenamefont {Fleury}(1968)}]{fleury_evidence_1968}%
  \BibitemOpen
  \bibfield  {author} {\bibinfo {author} {\bibfnamefont {P.~A.}\ \bibnamefont
  {Fleury}},\ }\bibfield  {title} {\bibinfo {title} {Evidence for
  {Magnon}-{Magnon} {Interactions} in {RbMnF$_3$}},\ }\href
  {https://doi.org/10.1103/PhysRevLett.21.151} {\bibfield  {journal} {\bibinfo
  {journal} {Physical Review Letters}\ }\textbf {\bibinfo {volume} {21}},\
  \bibinfo {pages} {151} (\bibinfo {year} {1968})}\BibitemShut {NoStop}%
\bibitem [{\citenamefont {Zhitomirsky}\ and\ \citenamefont
  {Chernyshev}(2013)}]{zhitomirsky_colloquium_2013}%
  \BibitemOpen
  \bibfield  {author} {\bibinfo {author} {\bibfnamefont {M.~E.}\ \bibnamefont
  {Zhitomirsky}}\ and\ \bibinfo {author} {\bibfnamefont {A.~L.}\ \bibnamefont
  {Chernyshev}},\ }\bibfield  {title} {\bibinfo {title} {Colloquium:
  {Spontaneous} magnon decays},\ }\href
  {https://doi.org/10.1103/RevModPhys.85.219} {\bibfield  {journal} {\bibinfo
  {journal} {Reviews of Modern Physics}\ }\textbf {\bibinfo {volume} {85}},\
  \bibinfo {pages} {219} (\bibinfo {year} {2013})}\BibitemShut {NoStop}%
\bibitem [{\citenamefont {Mentink}(2017)}]{mentink_manipulating_2017}%
  \BibitemOpen
  \bibfield  {author} {\bibinfo {author} {\bibfnamefont {J.~H.}\ \bibnamefont
  {Mentink}},\ }\bibfield  {title} {\bibinfo {title} {Manipulating magnetism by
  ultrafast control of the exchange interaction},\ }\href
  {https://doi.org/10.1088/1361-648X/aa8abf} {\bibfield  {journal} {\bibinfo
  {journal} {Journal of Physics: Condensed Matter}\ }\textbf {\bibinfo {volume}
  {29}},\ \bibinfo {pages} {453001} (\bibinfo {year} {2017})}\BibitemShut
  {NoStop}%
\bibitem [{\citenamefont {Batignani}\ \emph {et~al.}(2015)\citenamefont
  {Batignani}, \citenamefont {Bossini}, \citenamefont {Di~Palo}, \citenamefont
  {Ferrante}, \citenamefont {Pontecorvo}, \citenamefont {Cerullo},
  \citenamefont {Kimel},\ and\ \citenamefont
  {Scopigno}}]{batignani_probing_2015}%
  \BibitemOpen
  \bibfield  {author} {\bibinfo {author} {\bibfnamefont {G.}~\bibnamefont
  {Batignani}}, \bibinfo {author} {\bibfnamefont {D.}~\bibnamefont {Bossini}},
  \bibinfo {author} {\bibfnamefont {N.}~\bibnamefont {Di~Palo}}, \bibinfo
  {author} {\bibfnamefont {C.}~\bibnamefont {Ferrante}}, \bibinfo {author}
  {\bibfnamefont {E.}~\bibnamefont {Pontecorvo}}, \bibinfo {author}
  {\bibfnamefont {G.}~\bibnamefont {Cerullo}}, \bibinfo {author} {\bibfnamefont
  {A.}~\bibnamefont {Kimel}},\ and\ \bibinfo {author} {\bibfnamefont
  {T.}~\bibnamefont {Scopigno}},\ }\bibfield  {title} {\bibinfo {title}
  {Probing ultrafast photo-induced dynamics of the exchange energy in a
  {Heisenberg} antiferromagnet},\ }\href
  {https://doi.org/10.1038/nphoton.2015.121} {\bibfield  {journal} {\bibinfo
  {journal} {Nature Photonics}\ }\textbf {\bibinfo {volume} {9}},\ \bibinfo
  {pages} {506} (\bibinfo {year} {2015})}\BibitemShut {NoStop}%
\bibitem [{\citenamefont {Nova}\ \emph {et~al.}(2017)\citenamefont {Nova},
  \citenamefont {Cartella}, \citenamefont {Cantaluppi}, \citenamefont {Först},
  \citenamefont {Bossini}, \citenamefont {Mikhaylovskiy}, \citenamefont
  {Kimel}, \citenamefont {Merlin},\ and\ \citenamefont
  {Cavalleri}}]{nova_effective_2017}%
  \BibitemOpen
  \bibfield  {author} {\bibinfo {author} {\bibfnamefont {T.~F.}\ \bibnamefont
  {Nova}}, \bibinfo {author} {\bibfnamefont {A.}~\bibnamefont {Cartella}},
  \bibinfo {author} {\bibfnamefont {A.}~\bibnamefont {Cantaluppi}}, \bibinfo
  {author} {\bibfnamefont {M.}~\bibnamefont {Först}}, \bibinfo {author}
  {\bibfnamefont {D.}~\bibnamefont {Bossini}}, \bibinfo {author} {\bibfnamefont
  {R.~V.}\ \bibnamefont {Mikhaylovskiy}}, \bibinfo {author} {\bibfnamefont
  {A.~V.}\ \bibnamefont {Kimel}}, \bibinfo {author} {\bibfnamefont
  {R.}~\bibnamefont {Merlin}},\ and\ \bibinfo {author} {\bibfnamefont
  {A.}~\bibnamefont {Cavalleri}},\ }\bibfield  {title} {\bibinfo {title} {An
  effective magnetic field from optically driven phonons},\ }\href
  {https://doi.org/10.1038/nphys3925} {\bibfield  {journal} {\bibinfo
  {journal} {Nature Physics}\ }\textbf {\bibinfo {volume} {13}},\ \bibinfo
  {pages} {132} (\bibinfo {year} {2017})}\BibitemShut {NoStop}%
\bibitem [{\citenamefont {Czachor}(1995)}]{czachor_paramagnetic_1995}%
  \BibitemOpen
  \bibfield  {author} {\bibinfo {author} {\bibfnamefont {A.}~\bibnamefont
  {Czachor}},\ }\bibfield  {title} {\bibinfo {title} {Paramagnetic {Curie}
  temperature is an arithmetic average of the interspin coupling constants},\
  }\href {https://doi.org/10.1016/0304-8853(95)90014-4} {\bibfield  {journal}
  {\bibinfo  {journal} {Journal of Magnetism and Magnetic Materials}\ }\textbf
  {\bibinfo {volume} {139}},\ \bibinfo {pages} {355} (\bibinfo {year}
  {1995})}\BibitemShut {NoStop}%
\bibitem [{\citenamefont {Moriya}(1960)}]{moriya_anisotropic_1960}%
  \BibitemOpen
  \bibfield  {author} {\bibinfo {author} {\bibfnamefont {T.}~\bibnamefont
  {Moriya}},\ }\bibfield  {title} {\bibinfo {title} {Anisotropic
  {Superexchange} {Interaction} and {Weak} {Ferromagnetism}},\ }\href
  {https://doi.org/10.1103/PhysRev.120.91} {\bibfield  {journal} {\bibinfo
  {journal} {Physical Review}\ }\textbf {\bibinfo {volume} {120}},\ \bibinfo
  {pages} {91} (\bibinfo {year} {1960})}\BibitemShut {NoStop}%
\bibitem [{\citenamefont {Moskvin}(2019)}]{moskvin_dzyaloshinskiimoriya_2019}%
  \BibitemOpen
  \bibfield  {author} {\bibinfo {author} {\bibfnamefont {A.}~\bibnamefont
  {Moskvin}},\ }\bibfield  {title} {\bibinfo {title} {Dzyaloshinskii–{Moriya}
  {Coupling} in 3d {Insulators}},\ }\href
  {https://doi.org/10.3390/condmat4040084} {\bibfield  {journal} {\bibinfo
  {journal} {Condensed Matter}\ }\textbf {\bibinfo {volume} {4}},\ \bibinfo
  {pages} {84} (\bibinfo {year} {2019})}\BibitemShut {NoStop}%
\bibitem [{\citenamefont {Dmitrienko}\ \emph {et~al.}(2014)\citenamefont
  {Dmitrienko}, \citenamefont {Ovchinnikova}, \citenamefont {Collins},
  \citenamefont {Nisbet}, \citenamefont {Beutier}, \citenamefont {Kvashnin},
  \citenamefont {Mazurenko}, \citenamefont {Lichtenstein},\ and\ \citenamefont
  {Katsnelson}}]{dmitrienko_measuring_2014}%
  \BibitemOpen
  \bibfield  {author} {\bibinfo {author} {\bibfnamefont {V.~E.}\ \bibnamefont
  {Dmitrienko}}, \bibinfo {author} {\bibfnamefont {E.~N.}\ \bibnamefont
  {Ovchinnikova}}, \bibinfo {author} {\bibfnamefont {S.~P.}\ \bibnamefont
  {Collins}}, \bibinfo {author} {\bibfnamefont {G.}~\bibnamefont {Nisbet}},
  \bibinfo {author} {\bibfnamefont {G.}~\bibnamefont {Beutier}}, \bibinfo
  {author} {\bibfnamefont {Y.~O.}\ \bibnamefont {Kvashnin}}, \bibinfo {author}
  {\bibfnamefont {V.~V.}\ \bibnamefont {Mazurenko}}, \bibinfo {author}
  {\bibfnamefont {A.~I.}\ \bibnamefont {Lichtenstein}},\ and\ \bibinfo {author}
  {\bibfnamefont {M.~I.}\ \bibnamefont {Katsnelson}},\ }\bibfield  {title}
  {\bibinfo {title} {Measuring the {Dzyaloshinskii}–{Moriya} interaction in a
  weak ferromagnet},\ }\href {https://doi.org/10.1038/nphys2859} {\bibfield
  {journal} {\bibinfo  {journal} {Nature Physics}\ }\textbf {\bibinfo {volume}
  {10}},\ \bibinfo {pages} {202} (\bibinfo {year} {2014})}\BibitemShut
  {NoStop}%
\bibitem [{\citenamefont {Fogh}\ \emph {et~al.}(2019)\citenamefont {Fogh},
  \citenamefont {Zaharko}, \citenamefont {Schefer}, \citenamefont
  {Niedermayer}, \citenamefont {Holm-Dahlin}, \citenamefont {Sørensen},
  \citenamefont {Kristensen}, \citenamefont {Andersen}, \citenamefont {Vaknin},
  \citenamefont {Christensen},\ and\ \citenamefont
  {Toft-Petersen}}]{fogh_dzyaloshinskii-moriya_2019}%
  \BibitemOpen
  \bibfield  {author} {\bibinfo {author} {\bibfnamefont {E.}~\bibnamefont
  {Fogh}}, \bibinfo {author} {\bibfnamefont {O.}~\bibnamefont {Zaharko}},
  \bibinfo {author} {\bibfnamefont {J.}~\bibnamefont {Schefer}}, \bibinfo
  {author} {\bibfnamefont {C.}~\bibnamefont {Niedermayer}}, \bibinfo {author}
  {\bibfnamefont {S.}~\bibnamefont {Holm-Dahlin}}, \bibinfo {author}
  {\bibfnamefont {M.~K.}\ \bibnamefont {Sørensen}}, \bibinfo {author}
  {\bibfnamefont {A.~B.}\ \bibnamefont {Kristensen}}, \bibinfo {author}
  {\bibfnamefont {N.~H.}\ \bibnamefont {Andersen}}, \bibinfo {author}
  {\bibfnamefont {D.}~\bibnamefont {Vaknin}}, \bibinfo {author} {\bibfnamefont
  {N.~B.}\ \bibnamefont {Christensen}},\ and\ \bibinfo {author} {\bibfnamefont
  {R.}~\bibnamefont {Toft-Petersen}},\ }\bibfield  {title} {\bibinfo {title}
  {Dzyaloshinskii-{Moriya} interaction and the magnetic ground state in
  magnetoelectric {LiCoPO$_4$}},\ }\href
  {https://doi.org/10.1103/PhysRevB.99.104421} {\bibfield  {journal} {\bibinfo
  {journal} {Physical Review B}\ }\textbf {\bibinfo {volume} {99}},\ \bibinfo
  {pages} {104421} (\bibinfo {year} {2019})}\BibitemShut {NoStop}%
\bibitem [{\citenamefont {Sergienko}\ and\ \citenamefont
  {Dagotto}(2006)}]{sergienko_role_2006}%
  \BibitemOpen
  \bibfield  {author} {\bibinfo {author} {\bibfnamefont {I.~A.}\ \bibnamefont
  {Sergienko}}\ and\ \bibinfo {author} {\bibfnamefont {E.}~\bibnamefont
  {Dagotto}},\ }\bibfield  {title} {\bibinfo {title} {Role of the
  {Dzyaloshinskii}-{Moriya} interaction in multiferroic perovskites},\ }\href
  {https://doi.org/10.1103/PhysRevB.73.094434} {\bibfield  {journal} {\bibinfo
  {journal} {Physical Review B}\ }\textbf {\bibinfo {volume} {73}},\ \bibinfo
  {pages} {094434} (\bibinfo {year} {2006})}\BibitemShut {NoStop}%
\bibitem [{\citenamefont {Lemmens}\ \emph {et~al.}(2003)\citenamefont
  {Lemmens}, \citenamefont {Güntherodt},\ and\ \citenamefont
  {Gros}}]{lemmens_magnetic_2003}%
  \BibitemOpen
  \bibfield  {author} {\bibinfo {author} {\bibfnamefont {P.}~\bibnamefont
  {Lemmens}}, \bibinfo {author} {\bibfnamefont {G.}~\bibnamefont
  {Güntherodt}},\ and\ \bibinfo {author} {\bibfnamefont {C.}~\bibnamefont
  {Gros}},\ }\bibfield  {title} {\bibinfo {title} {Magnetic light scattering in
  low-dimensional quantum spin systems},\ }\href
  {https://doi.org/10.1016/S0370-1573(02)00321-6} {\bibfield  {journal}
  {\bibinfo  {journal} {Physics Reports}\ }\textbf {\bibinfo {volume} {375}},\
  \bibinfo {pages} {1} (\bibinfo {year} {2003})}\BibitemShut {NoStop}%
\bibitem [{\citenamefont {Jeandey}\ and\ \citenamefont
  {Nechtschein}(1980)}]{jeandey_TFF_1980}%
  \BibitemOpen
  \bibfield  {author} {\bibinfo {author} {\bibfnamefont {C.}~\bibnamefont
  {Jeandey}}\ and\ \bibinfo {author} {\bibfnamefont {M.}~\bibnamefont
  {Nechtschein}},\ }\bibfield  {title} {\bibinfo {title} {Inter- and
  intra-chain couplings in {TTF} {CuBDT} as determined from proton spin lattice
  relaxation time measurements},\ }\href
  {https://doi.org/10.1016/0304-8853(80)90884-7} {\bibfield  {journal}
  {\bibinfo  {journal} {Journal of Magnetism and Magnetic Materials}\ }\textbf
  {\bibinfo {volume} {15-18}},\ \bibinfo {pages} {1053} (\bibinfo {year}
  {1980})}\BibitemShut {NoStop}%
\bibitem [{\citenamefont {Nishi}\ \emph {et~al.}(1994)\citenamefont {Nishi},
  \citenamefont {Fujita},\ and\ \citenamefont {Akimitsu}}]{nishi_CuGeO_1994}%
  \BibitemOpen
  \bibfield  {author} {\bibinfo {author} {\bibfnamefont {M.}~\bibnamefont
  {Nishi}}, \bibinfo {author} {\bibfnamefont {O.}~\bibnamefont {Fujita}},\ and\
  \bibinfo {author} {\bibfnamefont {J.}~\bibnamefont {Akimitsu}},\ }\bibfield
  {title} {\bibinfo {title} {Neutron-scattering study on the spin-{Peierls}
  transition in a quasi-one-dimensional magnet {CuGeO$_3$}},\ }\href
  {https://doi.org/10.1103/PhysRevB.50.6508} {\bibfield  {journal} {\bibinfo
  {journal} {Physical Review B}\ }\textbf {\bibinfo {volume} {50}},\ \bibinfo
  {pages} {6508} (\bibinfo {year} {1994})}\BibitemShut {NoStop}%
\bibitem [{\citenamefont {Hirakawa}\ and\ \citenamefont
  {Kurogi}(1970)}]{hirakawa_KCuF_1970}%
  \BibitemOpen
  \bibfield  {author} {\bibinfo {author} {\bibfnamefont {K.}~\bibnamefont
  {Hirakawa}}\ and\ \bibinfo {author} {\bibfnamefont {Y.}~\bibnamefont
  {Kurogi}},\ }\bibfield  {title} {\bibinfo {title} {One-{Dimensional}
  {Antiferromagnetic} {Properties} of {KCuF$_3$}},\ }\href
  {https://doi.org/10.1143/PTPS.46.147} {\bibfield  {journal} {\bibinfo
  {journal} {Progress of Theoretical Physics Supplement}\ }\textbf {\bibinfo
  {volume} {46}},\ \bibinfo {pages} {147} (\bibinfo {year} {1970})}\BibitemShut
  {NoStop}%
\end{thebibliography}%

\end{document}